\documentclass[conference]{IEEEtran}

\usepackage{graphicx}
\usepackage{color}
\usepackage{algorithm}
\usepackage{algorithmicx}
\usepackage[lined,boxed,commentsnumbered]{algorithm2e}
\usepackage{epstopdf}
\begin{document}

\title{A GPU-accelerated Branch-and-Bound Algorithm for the Flow-Shop Scheduling Problem}

\author{ \IEEEauthorblockN{ N. Melab$^*$ , I. Chakroun $^*$, M. Mezmaz$^{**}$ and D.Tuyttens$^{**}$ } 

\IEEEauthorblockA{$^*$ Universit\'e Lille~1, LIFL/UMR CNRS 8022\\
59655 - Villeneuve d'Ascq cedex - France \\
Email: \{nouredine.melab, imen.chakroun\}@lifl.fr}
\IEEEauthorblockA{Mathematics and OR Department, University of Mons, Belgium\\
Email: \{mohand.mezmaz,daniel.tuyttens\}@umons.ac.be}

}

\maketitle              

\begin{abstract}

Branch-and-Bound (B\&B) algorithms are time-intensive tree-based exploration methods for solving 
to optimality combinatorial optimization problems. In this paper, we investigate the use 
of GPU computing as a major complementary way to speed up those methods. The focus is put on the bounding 
mechanism of B\&B algorithms, which is the most time consuming part of their exploration process. 
We propose a parallel B\&B algorithm based on a GPU-accelerated bounding model. The proposed approach 
concentrate on optimizing data access management to further improve the performance of the 
bounding mechanism which uses large and intermediate data sets that do not completely fit in GPU memory.
Extensive experiments of the contribution have been carried out on well-known FSP benchmarks 
using an Nvidia Tesla C2050 GPU card. We compared the obtained performances
to a single and a multi-threaded CPU-based execution. Accelerations up to $\times 100$ 
are achieved for large problem instances.

\end{abstract}

\vspace{0.5cm}


\section{Introduction}

Combinatorial optimization problems\footnote{An optimization problem consists in minimizing or maximizing a cost 
function. Without loss of generality, in this paper the minimization case is considered.} are NP-hard and CPU-time intensive in practice. 
Branch-and-Bound (B\&B) algorithms are efficient methods for solving to optimality those problems. Their execution 
consists in exploring a search space by dynamically building a tree whose root node is the original problem, the 
intermediate nodes are sub-problems, and the leaves are potential solution(s). B\&B proceeds in several iterations 
during which the best solution found so far (upper bound) is progressively improved. During the exploration, a bounding
mechanism, based on a lower bound function, is used to eliminate all the sub-problems (i.e. cut their corresponding sub-trees) 
that are not likely to lead to optimal solutions. Such powerful mechanism allows one to reduce significantly the size of the 
explored search space and thus its exploration time cost.

However, even if such mechanism is highly efficient it is not sufficient to deal with large size problem instances. Over the 
last decades, parallel computing has emerged as an attractive way to deal with larger instances. The design and implementation 
of parallel B\&B is strongly influenced by the computing platform. Many
contributions have been proposed for the design and implementation of parallel B\&B methods using Massively Parallel Processors 
(MPP)~\cite{Allen_1997}, Networks or Clusters of Workstations (NOWs or COWs)~\cite{Tschoke_1995} and Shared Memory or SMP 
machines~\cite{Casadoa_2008}. In this paper, we investigate the design of B\&B algorithms on Graphics Processing Units (GPU). 
In combinatorial optimization, GPU computing is successfuly used for meta-heuristics (near-optimal methods)~\cite{Luong_2012} but not yet for B\&B exact methods. 

Most of existing parallel B\&B algorithms are based on the parallel exploration of the search tree. Such parallel model is 
not suited to GPUs because the explored search tree is highly irregular. The best parallel model for B\&B on GPU is the parallel 
evaluation of the lower bound function (thread kernel) on pools of sub-problems (parallel bounding). Such model must be 
rethought at design as well as at implementation level taking into account at the same time the characteristics of GPU 
accelerators and those of the lower bound computation function. On the one hand, a GPU is a many-core co-processor device 
that provides a hierarchy of memories having different sizes and access latencies making data placement and sharing challenging. 
On the other hand, the lower bound computation function is generally problem-dependent. In this paper, the focus is on the 
Flow-Shop scheduling permutation Problem (FSP) (see Section~\ref{BB-FSP}). The lower bound function used in this work for 
FSP is that proposed in~\cite{SMJohnson_54} for two machines and generalized in~\cite{JKLenstra_et_al_78} to more than two machines. 
The implementation of such function makes use of six data structures of different sizes and access frequencies making data placement 
on GPU challenging. 

Preliminary experiments we have carried out on some Taillard's FSP instances~\cite{Taillard_Bench} have shown that the time spent by 
B\&B evaluating the lower bounds of the examined sub-problems is on average around $98.5\%$ of its total execution time. Such result 
illustrates the potential benefit of parallelizing the bounding operation. The major contribution of this paper consists in revisiting 
B\&B to allow efficient solving of large FSP instances on GPU. Having in mind the characteristics of both the lower bound function and 
the GPU device mentioned above, the challenge is to define a new approach for optimal mapping of the data structures of the lower bound 
function on the hierarchy of memories provided in the GPU device. A careful analysis is required of both the data structures (size and 
access frequency) and the GPU memories (size and access latency). 

The remainder of the paper is organized as follows: Section~\ref{BB} presents the B\&B algorithm applied to the permutation FSP, 
the associated lower bound used in this paper, and its implementation and complexity analysis. In Section~\ref{GPUBounding}, we
describe our GPU-based proposed approach for B\&B applied to FSP. Details are given on the parallel approach and memory access 
optimization. In Section~\ref{Experiments}, we report experimental results demonstrating the efficiency of our approach. In Section
~\ref{comparison} we compare the performance of the proposed approach to a multi-threaded CPU version of the B\&B. Finally, 
some conclusions and perspectives of this work are drawn in Section~\ref{Conclusion}.

\section{B\&B and Lower Bound for the Permutation FSP}
\label{BB}

\subsection{Parallel B\&B algorithms}
\label{P-BB}

Branch-and-Bound (B\&B) algorithms are based on an implicit enumeration of the solutions composing the search space associated to the problem to be tackled. 
The search space is explored by dynamically building a tree whose root node designates the original problem. The construction of the B\&B 
tree and its exploration are performed using four operators: {\it branching}, {\it bounding}, {\it selection} and {\it elimination}. The 
algorithm proceeds in several iterations during which the best solution found so far (upper bound) is progressively improved. The generated 
and not yet examined sub-problems are kept into a list initialized to the original problem. At each iteration, a sub-problem is selected 
from this list, according to some strategy (depth-first, best-first, $\ldots$), using the {\it selection operator}. 
The {\it branching operator} performs its decomposition into other sub-problems. The {\it bounding operator} calculates a lower 
bound of each generated sub-problem. Each sub-problem having a lower bound greater than the upper bound is eliminated using the 
{\it elimination operator}, this means that it will not be decomposed.

Existing parallel B\&B algorithms are based on three parallel models proposed in~\cite{BGendron_et_al_94}: parallel application
 of the operators on the generated sub-problems (Type~1), parallel building and exploration of a B\&B tree (Type~2), and
parallel (cooperative or independent) building and exploration of several B\&B trees (Type~3). We have later revisited these 
parallel approaches for large-scale computational grids~\cite{Djamai_2011} using Type~2 parallel model. Grid computing provides 
an impressive computing power to solve challenging instances in combinatorial optimization~\cite{Mezmaz_2007}. However, 
computational grids providing a huge amount of resources are not easily available and accessible for any user. Recently, 
Graphics Processing Units (GPU accelerators) have emerged as a new popular support for massively parallel computing. Such 
resources supply a great computing power, are energy-efficient and unlike grids they are highly available every where: laptops, 
desktops, clusters, etc. In the following, we revisit the Type~1 parallel model on GPU for solving Flow-Shop problems.

\subsection{B\&B for the permutation FSP}
\label{BB-FSP}
%~\cite{Lenstra78}

The general FSP~\cite{JKLenstra_et_al_78} consists in scheduling a pool of $n$ jobs on a set of $m$ machines such that each of 
the jobs $J_1$, $J_2$, \ldots, $J_n$ has to be processed on the machines $M_1$, $M_2$, \ldots, $M_m$ in that order. Job $J_i$ (i = 1, 2, \ldots, n) 
consists therefore of a sequence of $m$ operations $O_{i1}$, $O_{i2}$,  \ldots $O_{im}$; $O_{ik}$ being the processing of $J_i$ on $M_k$ 
during an uninterrupted processing time $p_{ik}$. $M_k$ (k = 1, 2, \ldots, m) can handle at most one job at a time. The objective is to 
find a processing order on each $M_k$ such that the time (makespan) required to complete all jobs is minimized. If the problem is 
restricted to the minimization over all permutation schedules, meaning with the same processing order on each machine, the resulting 
problem is called the permutation Flow-Shop problem, which is the focus of this work. In the remainder of this paper, FSP designates a permutation FSP. 

\begin{figure}
  \begin{center}
\includegraphics[width=9cm]{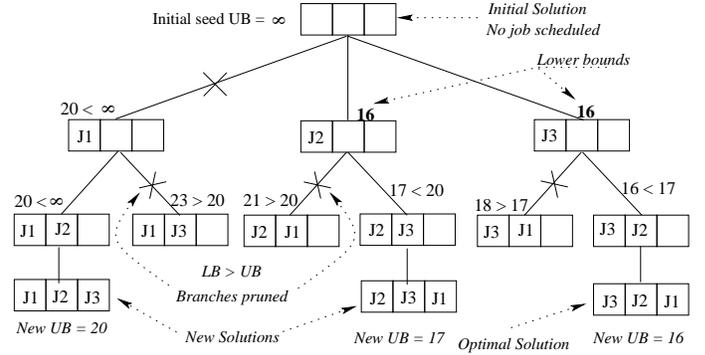}%
\caption{The search tree generated and explored by a B\&B algorithm for solving an FSP with 3 jobs. Nodes with a lower bound (LB) 
greater (resp. lower or equal) than the upper bound (UB) are pruned (resp. decomposed or branched).}
\label{BBExample}
 \end{center}
\end{figure}

For $m=2$, an optimal schedule can be found in $O(n.logn)$ steps using Johnson's algorithm~\cite{SMJohnson_54}. For $m \geq 3$, the 
problem has been shown to be NP-hard~\cite{MRGary_et_al_79}. Due to such complexity the enumerative solution approach provided in B\&B 
algorithms is well-suited to solve the problem to optimality.
As illustrated (for $n=3$) in Figure~\ref{BBExample}, the B\&B enumeration scheme is based on a search tree whose root node contains 
the original problem (empty schedule).

The decomposition of this problem generates $n$ sons, each of them designates a sub-problem. The son number $i$ represents the sub-problem 
in which job $J_i$ is scheduled first on all machines. The recursive application of the decomposition operator on the generated sub-problems 
allows to develop the search tree. The number
of potential schedules (permutations) is $n!$, which is highly
exorbitant for large problem instances such as $200 \times 20$
($200!$ schedules!) Taillard's ones~\cite{Taillard_Bench}. There are
two major powerful ways to speed up the exploration of large search
trees. The first way consists in using an efficient bounding
operator. Applied to a sub-problem, such operator associates a value
to its corresponding tree node using a lower bound function. As
illustrated in Figure~\ref{BBExample}, the sub-problem is not
decomposed (and its tree node is pruned) if its lower bound value is
greater than the cost of the best schedule found so far (called the
upper bound) during the exploration of the search tree. The second
way is to use massively parallel computing based on the three
parallelism types presented in Section~\ref{P-BB}. We recall
that the focus of this paper is only on {\it Type~1} i.e. the
parallel evaluation of the lower bound on a pool of sub-problems.

\subsection{Lower Bound for FSP}
As quoted above, the objective (cost function) of FSP considered in
this paper is the makespan $C_{max}$, which represents the
completion time of the last scheduled job on the last machine. Given
a sub-problem (partial schedule) $\pi = \pi(1), \pi(2), \ldots, \pi(l)$
indicating that $J_{\pi(i)}$ occupies the $i^{th}$ position on each
machine for $i=1, \ldots, l$. The sub-problem consists to find the
optimal schedule of the $n-l$ remaining unscheduled jobs. Before
solving such sub-problem, it is checked either or not the optimal
solution of the original problem could be the optimal solution of
this sub-problem. In other words, it is checked either the
optimal solution of the original problem is probably in the sub-tree
search space associated to that sub-problem or not. This is the role of the
bounding operator which uses a lower bound function to prune nodes
and the sub-trees they are root of. Indeed, if the lower bound value
of the sub-problem is greater than the upper bound found so far the
sub-problem is not decomposed/branched because it is sure that
the optimal solution is not located in its sub-tree search space.
This allows to significantly reduce the exploration time of the B\&B
tree. Therefore, the efficiency of a B\&B algorithm depends strongly
on the quality of its lower bound function. In this paper, we use the lower bound proposed by Lenstra {\it et al.}~\cite{JKLenstra_et_al_78} for FSP, based on the
Johnson's algorithm~\cite{SMJohnson_54}.

\subsection{Complexity analysis and implementation}
\label{MemComplex}

For an efficient implementation of the lower bound LB algorithm,
six data structures are required: the  matrix $PTM$ of the
processing times of the jobs, the matrix of lags $LM$, the Johnson's
matrix $JM$, the matrix $RM$ of the earliest starting times of jobs, the matrix
$QM$ of their lowest latency times and the matrix $MM$ containing the couples of machines. In the $LB$ expression, the computation of 
the term $P_{Ja}^*(\jmath,M_k,M_l)$ requires the calculation of the lag of each remaining job to be scheduled on the couple $(M_k,M_l)$ of machines using its processing
times on these machines (Johnson's rule with lags). Such computation
is repeated for each couple $(M_k,M_l)$ of machines with $1 \leq k,l
\leq m$ and $k<l$. To avoid the repetitive computation of the lags,
they are computed once at the beginning of the algorithm and stored
in the matrix $LM$. The dimension of $LM$ is $n \times \frac{m
\times (m-1)}{2}$, where $n$ and $m$ are respectively the number of
jobs to be scheduled and $m$ the number of machines. $LM$ is
accessed $n' \times \frac{m \times (m-1)}{2}$ times, $n'$ being the
number of remaining jobs to be scheduled in the sub-problem for
which the lower bound is being calculated. The processing times of
all the jobs on all the machines are stored in the matrix $PTM$.
This matrix has a dimension of $n \times m$ and is accessed $n' \times m \times (m-1)$ times.\\

Table~\ref{tab:MemComplex}, is highly needed to understand the proposed data placement approach. The columns of Table~\ref{tab:MemComplex} 
represent respectively the name of the data structure, its size and the number of times it is accessed. Figure~\ref{LB-pseudocode} shows 
the pseudo-code implementing the $LB$ lower bound function illustrating the access to the six data structures.

\begin{table}
  \centering
\begin{tabular}{|c|c|c|}
\hline
 % after \\: \hline or \cline{col1-col2} \cline{col3-col4} ...
 \textbf{Matrix} & \textbf{Size} & \textbf{Number of accesses} \\
 \hline
 \hline
   PTM &  $n \times m$ & $n' \times m \times (m-1)$ \\
 \hline
   LM & $n \times \frac{m \times (m-1)}{2}$ & $n' \times \frac{m \times (m-1)}{2}$ \\
 \hline
   JM & $n \times \frac{m \times (m-1)}{2}$ & $n \times \frac{m \times (m-1)}{2}$ \\
 \hline
   RM &  $m$ & $m \times (m-1)$ \\
 \hline
   QM &  $m$ & $\frac{m \times (m-1)}{2}$ \\
 \hline
   MM &  $m \times (m-1)$ & $m \times (m-1)$ \\
 \hline
\end{tabular}
\vspace{0.3cm}
 \caption{The different data structures of the $LB$ algorithm and their associated complexities in memory size and numbers of accesses. 
The parameters $n$, $m$ and $n'$ designate resp. the total number of jobs, the total number of machines and the number of remaining 
jobs to be scheduled for the sub-problems the lower bound is being computed.}
\label{tab:MemComplex}
\end{table}

\begin{figure}[t]
{\small
$_{(01)}$ \textbf{int} \emph{computeLB}()\{ \\
$_{(02)}$\hspace{0.5cm}LB=maxInteger;\\
$_{(03)}$\hspace{0.5cm}\textbf{for} (index=0;index$<\frac{m\times(m-1)}{2}$;index++)\{ \\
$_{(04)}$\hspace{1cm}M1=\textbf{MM}[index][0]; \\
$_{(05)}$\hspace{1cm}M2=\textbf{MM}[index][1]; \\
$_{(06)}$\hspace{1cm}timeOnM1=$\min\limits_{0\leq j \leq n}$(\textbf{RM}[M1][j]);\\
$_{(07)}$\hspace{1cm}timeOnM2=$\min\limits_{0\leq j \leq n}$(\textbf{RM}[M2][j]);\footnote{Only on access to RM because the minimum values are computed on CPU.}\\
%$_{(05)}$\hspace{1cm}\emph{computeStartingTimes}(); // Access to RM \\
$_{(08)}$\hspace{1cm}\textbf{for} (i=0;i$<$n;i++)\{\\
$_{(09)}$\hspace{1.5cm}job=\textbf{JM}[i][index];\\
$_{(10)}$\hspace{1.5cm}\textbf{if} (\emph{job not yet scheduled})\{\\
$_{(11)}$\hspace{2cm} timeOnM1=timeOnM1+PTM[job][M1];\\
$_{(12)}$\hspace{2cm}\textbf{if} (timeOnM2$>$timeOnM1+\textbf{LM}[job][index])$^{(*)}$\\
$_{(13)}$\hspace{2.5cm}timeOnM2+=\textbf{PTM}[job][M2];\\
$_{(14)}$\hspace{2cm}\textbf{else}\\
$_{(15)}$\hspace{2.5cm}timeOnM2=timeOnM1+\textbf{LM}[job][index]+\\
.		\hspace{6cm}\textbf{PTM}[job][M2];\\
$_{(16)}$\hspace{1.5cm}\}\\
%$_{(08)}$\hspace{1.5cm}.\\
%$_{(09)}$\hspace{1.5cm}.\\
$_{(17)}$\hspace{1cm}\}\\
%$_{(11)}$\}
%$_{(06)}$\hspace{1cm}\emph{computeCMaxWithLags}(index);\\%,PTM,LM,JM);\\
%$_{(07)}$\hspace{1cm}\emph{computeLatency}(); ~~~~~// Access to QM\\
$_{(18)}$\hspace{1cm}timeOnM2+=$\min\limits_{0\leq j \leq n}$(\textbf{QM}[M2][j]);\\
$_{(19)}$\hspace{1cm}LB=max(timeOnM2,LB);\\
%$_{(09)}$\hspace{1cm}.\\
$_{(20)}$\hspace{0.5cm}\}\\
%$_{(11)}$\hspace{0.5cm}.\\
$_{(21)}$\hspace{0.5cm}\textbf{return} LB;\\
$_{(22)}$\}
} \caption{Pseudo-code implementing the LB function}
\label{LB-pseudocode} 
\end{figure}

\section{GPU-based B\&B for FSP - A new Approach}
\label{GPUBounding} 
As said previously, the time complexity of the Johnson algorithm for two machines is $O(n. log n)$. Therefore, the time complexity of
the lower bound $LB$ for $m$ machines is $O(m^2.n. log n)$. The computation of the lower bound is consequently time intensive especially 
for problem instances for which $m$ is high. In order to evaluate experimentally its CPU time cost, we have implemented this lower bound 
and experimented it using the most time-consuming Taillard's instances~\cite{Taillard_Bench} i.e. having $m=20$. The results show that the 
time spent by the B\&B evaluating the lower bounds of the examined sub-problems is on average around $98.5\%$ of its total execution time. 
Such result demonstrates that the bounding must be parallelized i.e. the $LB$
lower bound function must be applied in parallel to each sub-problem composing the pool of sub-problems currently examined. In the following, 
we present a new GPU-based approach for the parallel evaluation of the lower bound in B\&B algorithms. We first present the parallel GPU-based
approach. Then, we show how our approach maps the different data structures on the memory hierarchy of the GPU device taking into account 
the characteristics of the data structures presented in Table~\ref{tab:MemComplex} and those of the different GPU memories (size and access latency). 

\subsection{The GPU-based parallel evaluation of LB}
The GPU-accelerated approach is based on the GPGPU (CUDA or OpenCL) parallel paradigm according to which the programmer writes a serial program 
that calls parallel kernels (simple functions or full programs). A kernel executes in parallel across a set of parallel threads. The programmer 
organizes these threads into a hierarchy of grids of thread blocks. A thread block is a set of concurrent threads that can cooperate through 
barrier synchronization and shared access to a memory space private to the block. A grid is a set of thread blocks that may be executed 
independently in parallel. When invoking a kernel, the programmer specifies the execution configuration. Such configuration includes the 
number of threads per block and the number of blocks making up the grid.

\begin{figure}
  \begin{center}
\includegraphics[width=9cm]{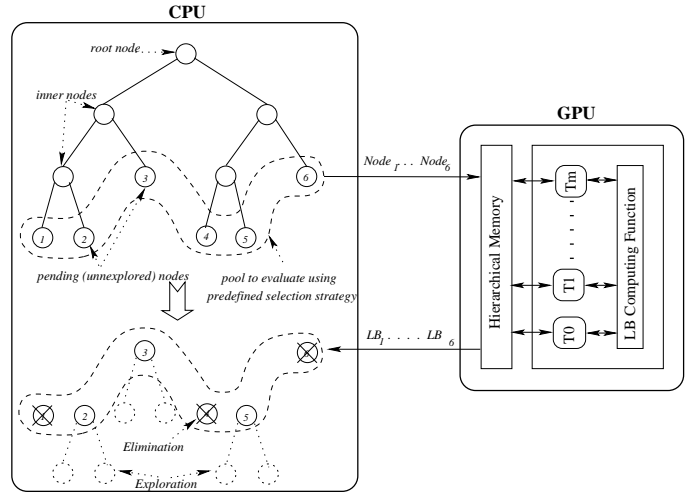}
\caption{GPU evaluation of sub-problems: generation on CPU and evaluation on GPU.}
\label{LB_GPUeval}
  \end{center}
\end{figure}

In our proposed GPU-based approach, the generation of the sub-problems (elimination, selection and branching operations) to be solved is 
performed on CPU and the evaluation of their lower bounds (bounding operation) is executed on the GPU device. As illustrated in Figure~\ref{LB_GPUeval}, 
the pool of sub-problems generated on CPU (and selected according to the well-know best-first strategy) is off-loaded to the GPU device to be evaluated 
by a pool of threads partitioned into blocks. Each thread applies the lower bound function (kernel) to one sub-problem. Once the evaluation is 
completed, the lower bound values of the different sub-problems are returned back to the CPU to be used by the elimination operator to decide 
either to be pruned or to be decomposed. The process is iterated until the exploration is completed and the optimal solution is found.

\subsection{Data access optimization}
\label{DataAccessOpt}
During their execution, threads may access data from multiple memory spaces having different sizes and access latencies. At the thread-level, 
each thread has its own allocated registers and a private local memory. CUDA~\cite{NvidiaCudaGuide} uses this local memory for thread-private 
variables that do not fit in the threads registers, as well as for stack frames and register spilling. At the thread block-level, each thread 
block has a shared memory visible to all its associated threads. At the grid-level, all threads have access to the same global memory. Texture 
and constant cached memories are two other memories accessible by all threads. The data access optimization challenge is to find the best 
mapping of the data structures of the application at hand (different sizes and access frequencies) and the GPU hierarchy of memories (different 
sizes and access latencies). For instance, the global memory is large in size but has a high access latency. On the contrary, shared memory is 
smaller in size but has a lower access latency.

For B\&B applied to FSP, threads of the same block perform concurrent accesses to the six data structures of the problem when they execute the 
LB lower bound function. To optimize the performance of such application, the best mapping of the data structures is to copy them on the shared 
memory of the GPU device. However, for large problem instances all of the data structures do not fit into the shared memory which size is limited
and depends on the GPU hardware configuration. The challenge is therefore to decide which data structure must be put in the shared memory to get 
the best performance. The answer is given in the next section according to the complexity analysis presented in Table~\ref{tab:MemComplex} and 
the underlying GPU configuration of our experiments.

\section{Experiments}
\label{Experiments} 

To evaluate the performance of our GPU-based B\&B algorithm and parallel bounding approach, 
we have considered the largest Taillard's FSP benchmarks proposed in~\cite{Taillard_Bench}, 
except those with $500$ jobs because they do not fit in the memory of the CPU. The different 
instances are designated by $n \times m$, where $n$ and $m$ represent respectively the number of jobs (between $20$ and $200$) 
to be scheduled and the number of machines ($20$) target of the scheduling. The GPU-based B\&B has been implemented using C-CUDA~4.0, 
and compiled using $nvcc$. The experiments have been carried out on an Intel Xeon E5520 paired with a GPU device. 
E5520 is 64-bit and composed of two quad-core chips, and has a clock speed of 2.27GHz. The GPU device is an Nvidia 
Tesla C2050 with $448$ CUDA cores ($14$ multiprocessors with $32$ cores each), a clock speed of 1.15GHz, a 2.8GB global memory, 
a configurable shared memory (16 KB or 48 KB) and a warp size of 32.

In the following, an experimental study is presented with the objective to evaluate the performance 
impact of the GPU-based parallel evaluation of the lower bound, and the data access optimization. 
For each, we present the objectives of the experiments and report the obtained results. Two parameters are considered: 
the problem instances ($n \times m$) (as rows in the tables and x-axis in the graphics) and the size of the pool of 
sub-problems to be evaluated (as columns in the tables and x-axis in the graphics). The first parameter gives information 
on the granularity of the thread computations. As the complexity of the computation of the lower bound is $O(m^2.n.logn)$, 
for large problem instances (i.e. large values of $n$ and $m$) the grain size of the kernel executed by each thread is much higher. 
Moreover, the first parameter gives information on the size of the data structures to be mapped on the GPU memories. This is highly 
important for the study of the data access optimization approach. The second parameter is designated in the different 
experimental results by $pool\ size\ (block\ size\ \times\ number\ of\ threads/block)$. This parameter is useful to get 
information on the time cost of the data transfer between CPU and GPU and on the total number of threads to be triggered on GPU.

For each pair of values associated to the two parameters, each table/graphics reports the corresponding parallel efficiency. 
Since the used instances are very hard to solve (optimal solutions for many of these instances are still not known), 
we used the approach defined in \cite{Mezmaz_2007} to run experiments. Employing this method 
allows to obtain a random list $L$ of subproblems such as the resolution of $L$ lasts $Tcpu$ minutes with a sequential B\&B. 
To ensure that the subproblems explored by the GPU and CPU B\&B versions are exactly the same, we initialize the pool 
of our GPU-based B\&B with the same list $L$ of subproblems used in the sequential version. 
If we suppose the resolution of the GPU-based B\&B last $T{gpu}$ minutes, the parallel efficiency would 
be the ratio $Tcpu/Tgpu$: the execution time of the serial B\&B on a single CPU core (without GPU)
over the execution time of our GPU-based B\&B on a CPU core coupled with a GPU device.

\subsection{Performance impact of GPU-based parallelism}

First, the objective of the experimental study presented in this section is to demonstrate that our GPU-based B\&B allows one 
to significantly accelerate the resolution process whatever is the FSP instance. However, the best achieved acceleration 
depends strongly on the problem instance being solved and the size of the pool of sub-problems considered at execution. 
The second objective is therefore to exhibit the behavior of the GPU acceleration according to the tackled problem instance 
and the considered pool size. More exactly, the goal is to find for each problem instance the best pool size required 
to maximize the benefit taken from the use of the GPU device.

\begin{table*}
  \centering
  \footnotesize
 \begin{tabular}{|r|r|r|r|r|r|r|r|}
    \hline
Problem instance &4096 & 8192 & 16384 & 32768 & 65536  & 131072 & 262144\\
& $16 \times $256 & $32 \times $256 & $64 \times $256 & $128 \times $256 & $256 \times $256 & $512 \times $256 & $1024 \times $256\\
    \hline
    \hline
$200 \times $20&46,63&60,88&63,80&67,51&73,47&75,94&\textbf{77,46}\\
    \hline
$100 \times $20&45,35&58,49&60,15&62,75&66,49&66,64&\textbf{67,01}\\
    \hline
$50 \times $20&44,39&\textbf{58,30}&57,72&57,68&57,37&57,01&56,42\\
    \hline
$20 \times $20&41,71&\textbf{50,28}&49,19&45,90&42,03&41,80&41,65\\
    \hline
    \hline
Average Speedup&44,52&56,99&57,72&58,46&59,84&60,35&60,64\\
    \hline
    \hline
 \end{tabular}
  \caption{Parallel efficiency for different problem instances and pool sizes. 
All the matrices $JM$, $PTM$, $LM$, $RM$, $QM$ and $MM$ are located in the GPU global memory.}
\label{tab:ParaGPU}
\end{table*}

The results reported in Table~\ref{tab:ParaGPU} are obtained without any data access optimization. 
The six matrices are generated on CPU and then copied to the GPU global memory. The size of the thread 
blocks is experimentally fixed to $256$ threads. Average accelerations of $\times 44,52$ to $\times 60,64$ 
and picked at $\times 77,46$ are achieved. In addition, the improvement of the parallel efficiency from a pool 
size of $4096\ (16 \times 256)$ to $8192\ (32 \times 256)$ is significant. The reason is that the number of blocks ($16$) 
for the first pool size is not sufficient to get a better acceleration. Indeed, it is known that the number of blocks 
must be fixed at least to the double ($14 \times 2 = 28$ for the C2050 GPU card) of the number of multi-processors of 
the target GPU device. Furthermore, for $50 \times 20$ and $20 \times 20$ problem instances the best parallel 
efficiency is achieved for a pool size of $8192$. For larger instances i.e. $100 \times 20$ and $200 \times 20$, 
it is obtained with a pool size of $262144$. These two pool size values correspond exactly to the two sizes of 
the pool for which the best ratio between lower bound evaluation time on CPU of the pool and its total communication time from CPU to GPU and from GPU to CPU.  

\subsection{Data access optimization}
\label{access}

The objective is here to find the best mapping of the six data structures of the lower bound LB kernel on the 
memories of the GPU device. As quoted in Section~\ref{DataAccessOpt}, such mapping depends on the sizes and 
access latencies/frequencies of these data structures and the GPU memories. The focus is put on the shared 
memory which is a key enabler for many high-performance CUDA applications. We also take care of adequately 
using the global memory by judiciously configuring the L1 cache that greatly enables improving performance over 
direct access to global memory. Indeed, the GPU device we are using in our experiments is a C2050 Tesla 
(see \ref{Experiments}) which a device based on the NVIDIA Fermi architecture. In the Fermi architecture,
each multiprocessor of the GPU device is provided with a $64$ KB local storage that can be configurable 
into shared memory and L1 cache. For this reason and in order to achieve further performances,
we divided the $64$ KB memory according to the scenario we are experimenting. For the scenario were the data structures
are put on the shared memory the $64$ KB of available storage are split on $48$ KB for shared memory and $16$ KB for L1 cache.
For the scenario where the data sets are put on global memory we used $16$ KB for shared memory and $48$ KB for L1 cache. 

As far as the data structures of the lower bound function are concerned, their complexities in terms of size and access frequency 
are reported in Table~\ref{tab:MemComplex} (see Section~\ref{MemComplex}). 
According to Table~\ref{tab:MemComplex}, $RM$, $QM$ and $MM$ have a small size, so their storage in the shared 
memory allows a very poor performance improvement. Therefore, whatever is the memory to which they are off-loaded, 
the performance impact is not significant. However, for large FSP instances (with $n=200$), the total amount of memory 
required to store the other data structures i.e. $JM$ and $LM$ ($38$KB each) and $PTM$ ($4$KB) is $80$KB, which is 
greater than the available shared memory space ($48$KB). Therefore, only two of them can be put in the shared memory. 
$LM$ has a double memory size than $JM$, and its access frequency is much lower, so it is better to map $JM$ on the shared memory. 
Furthermore, $PTM$ has the same access frequency than $JM$ but requires less memory space. Consequently, the focus is 
put on the study of the performance impact of the placement of $JM$ and $PTM$ on the shared memory. $PTM$ and $JM$ are 
stored in shared memory and all others are placed on global memory. 

\begin{table*}
  \centering
  \footnotesize
  \begin{tabular}{|r|r|r|r|r|r|r|r|}
    \hline
Problem instance &4096 & 8192 & 16384 & 32768 & 65536  & 131072 & 262144\\
& $16 \times $256 & $32 \times $256 & $64 \times $256 & $128 \times $256 & $256 \times $256 & $512 \times $256 & $1024 \times $256\\
    \hline
    \hline
$200 \times $20&66,13&87,34&88,861&95,23&98,83&99,89&\textbf{100,48}\\
    \hline
$100 \times $20&65,85&86,33&87,60&89,18&91,41&92,02&\textbf{92,39}\\
    \hline
$50 \times $20&64,91&\textbf{81,50}&78,02&74,16&73,83&73,25&72,71\\
    \hline
$20 \times $20&53,64&\textbf{61,47}&59,55&51,39&47,40&46,53&46,37\\
     \hline
    \hline
Average Speedup&62,63&79,16&78,51&77,49&77,87&77,92&77,99\\
    \hline
    \hline
  \end{tabular}
 \caption{Parallel efficiency for different FSP instances and pool sizes obtained with data access optimization. 
$PTM$ and $JM$ are placed together in shared memory and all others are placed in global memory.}
\label{tab:JM-PTM-on-SM}
\end{table*}

Table~\ref{tab:JM-PTM-on-SM} reports the behavior of the parallel efficiency averaged on the different problem 
instances (sizes) as a function of the pool size. The table 
shows that the parallel efficiency grows on average with the growing of the pool size in the same way as in 
Table~\ref{tab:ParaGPU}. For instance, for the largest problem instance and pool size, the parallel efficiency 
grows up to from $\times 77,46$ ($PTM$ and $JM$ in global memory) to $\times 100,48$ ($PTM$ and $JM$ in shared memory) ($23\%$).  

\begin{figure}
  \begin{center}
\includegraphics[width=9cm]{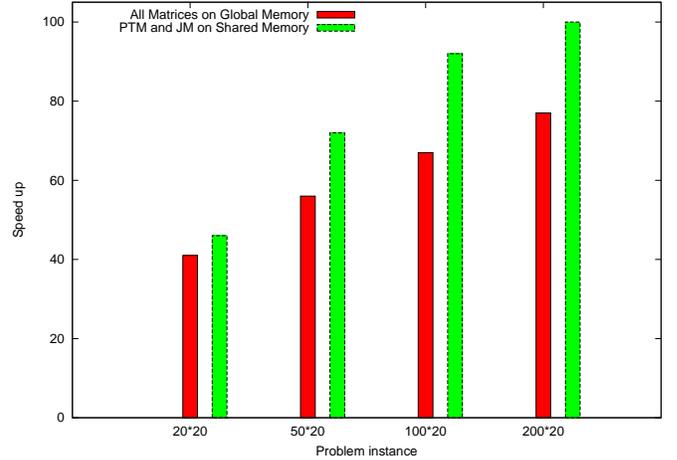}
\caption{Average parallel efficiency for different problem instances: $PTM$ and $JM$ are put together in the shared memory, the pool size is fixed to $1024 \times 256$.}
\label{fig:JM-PTM-Placement-Instances}
  \end{center}
\end{figure}

Figure~\ref{fig:JM-PTM-Placement-Instances} depicts the behavior of the parallel efficiency for the different 
problem instances (sizes). The pool size is fixed to $262144\ (1024 \times 256)$. According to the graphics, 
first, the efficiency is improved for all instances and the improvement is more significant for large problem 
instances. Second, the behavior of the efficiency improvement is not the same if shared memory is used or not. 
Indeed, according to the CUDA GPU occupancy calculator the size of the shared memory occupied by the data 
structures limits the number of active thread warps to $32$ for $20 \times 20$ and $50 \times 20$ problem 
instances, and to $16$ for $100 \times 20$ and $200 \times 20$ problem instances. When only global memory 
is used, the improvement is linear and the slop remains the same as the number of active thread warps 
remains the same ($32$) whatever is the problem instance. The only limiting factor of the active 
thread warps is the number of registers which is $26$ in our case. In this case, the size of the occupied shared 
memory is lower and is not a limiting factor for the occupancy or number of active threads. On the other hand, 
when shared memory is used the slope of the efficiency improvement is much higher from $20 \times 20$ to $50 \times 20$ 
(small data structures) than from $100 \times 20$ to $200 \times 20$ (large data structures). The reason is that according 
to CUDA GPU occupancy calculator in addition to the number of registers the size of the occupied shared memory is also a 
limiting factor of thread occupancy and thus parallel efficiency.   

\section{Performances comparison with a multi-threaded parallel B\&B algorithms}
\label{comparison}

With the advent of multi-core processors and their
promised enhancement in software development performances,
the use of multi-core processors for designing parallel algorithms 
become highly widespread. Unlike distributed computing systems, one of the 
advantages of multi-core systems is the possibility to parallelize the algorithm using threads instead of processes.
Indded, while processes in the same machine have their own virtual memory, 
threads of a process share the same virtual memory 
which significantly impact the performances.

Several implementation of a multi-threaded B\&B have been proposed in previous research works \cite{Barreto,Casadoa_2008,paulavicius,sanjuan}. These
multi-threaded B\&B algorithms can be classified into two categories: low and high-level. In a low-level multi-threaded B\&B, a
low-level thread model such as POSIX Threads is used \cite{Nichols_1996,Casadoa_2008} while in a high-level multi-threaded
B\&B a high-level thread model such as OpenMP \cite{Chapman_2007} is used.

In order to further evaluate the performances of the proposed GPU-based B\&B algorithm,
we compare it to a low-level multi-threaded B\&B \cite{Casadoa_2008} designed on top of a multi-core system,
using the POSIX Threads library. 

In order to perform a fair comparison with the obtained 
results of our GPU-based approach, the used multi core system must have the same computational power
in term of theoretical peak of floating-point operations per second. The floating-point operations per second (FLOPS) is a common measure of a 
computer's performance, especially in fields of scientific calculations. Indeed, FLOPS is a good indicator to measure performance 
on digital signal processing, scientific simulations, etc. It is particularly used in supercomputer ratings, like TOP500 \cite{TOP500}.

\begin{table*}
\setlength{\tabcolsep}{0.2cm}
\renewcommand{\arraystretch}{1.2}
  \centering
  \footnotesize
 \begin{tabular}{|c|c|c|c|c|c|}
    \hline
Number of B\&B Threads & 3 & 5 	& 7 & 9 & 11\\
Theoretical Peak of GFLOPS & 230.4 & 384 & 537.6 & 691.2 & 844.8\\
    \hline
    \hline
$200 \times $20& 4,03 & 6,98 &  8,76 & 9,04 & 9,32\\

    \hline
$100 \times $20& 4,27 & 7,08 &  8,82 & 9,39 & 9,85\\ 					

    \hline
$50 \times $20& 4,38 & 7,27 &  9,06 & 9,64 & 10,17\\
    \hline
$20 \times $20& 4,43 & 7,35 &  9,22 & 10,04 & 10,85\\
    \hline
    \hline
  \end{tabular}
\vspace{0.3cm}
  \caption{Parallel efficiency for different problem instances using the multi-threaded based B\&B.}
\label{tab:MULTI}
\end{table*}

As quoted in \ref{Experiments}, the experiments have been carried out on an Nvidia 
Tesla C2050. According to its constructor NVIDIA \cite{NVIDIA_C2050}, the theoretical double precision floating-point 
performance peak of this GPU device is about 515 GFLOPS. For the multi-threaded version of the B\&B we have carried 
out experimentation on an Intel Core i7-970 Processor which is 64-bit and composed of six physical cores and 12 
threads \cite{INTEL_I7_2} having each a theoretical double precision floating-point performance peak of 76.8 GFLOPS \cite{INTEL_I7}.

Table \ref{tab:MULTI} reports the speedup of the parallel multi-threaded B\&B averaged on the different problem 
instances (sizes). The columns correspond to the number of parallel running B\&B process
and the corresponding theoretical peak of GFLOPS. The rows correspond to the problem instances defined by 
(Number of jobs $\times$ Number of machines). The same experimental protocol as the for GPU computation is used (see section \ref{Experiments}).
The reported speedups are calculated relatively to a serial B\&B on a single CPU core.
Results shows that the parallel efficiency grows on average with the growing of the number of 
computing core used. However, the improvement is not linear and the slop decrease as long as the number of the used computing core
raises. This behavior might be due to the operating system which handles additional page faults and context switches
when the number of threads increases. 

\begin{figure}
  \begin{center}
\includegraphics[width=9cm]{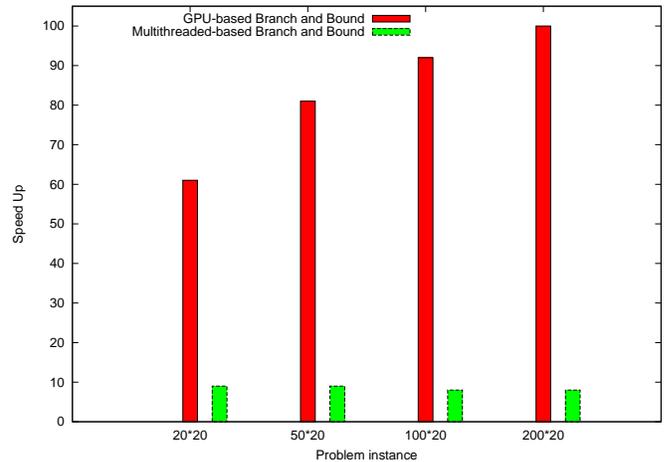}
\caption{Comparison between the average parallel efficiency for different problem instances obtained with a GPU and
a multithreaded-based B\&B for a same computational power (500 GFLOPs).}
\label{fig:MULTITHREADED}
  \end{center}
\end{figure}

Figure \ref{fig:MULTITHREADED} shows the comparison between the 
obtained speedups with our GPU-based B\&B and the multithreaded-based B\&B. The speedups are calculated
relatively to the same sequentiel version of the B\&B algorithm. For a same computational power, our 
approach for designing B\&B algorithms on top of GPU accelerators is much more efficient than the multi-threaded
B\&B whatever the instance is. Indeed, for a computational power around 500 GFLOPS, 
the acceleration calculated when using the GPU-based B\&B for the instances 20 jobs over 20 machines is $\times$61,47. 
For the same category of instances (20 jobs over 20 machines) and a same computational power of 500 GFLOPS which 
corresponds to 7 CPU computing cores for the Intel Core i7-970 Processor, the speedup over a sequential version of the multi-threaded based B\&B is $\times$9,22.
Results show also that parallel efficiency for the GPU-based approach increases with the size of the problem being tackled
while it is almost the same for the multi-threaded based algorithm. This is due to the complexity of the computation of the 
lower bound which is $O(m^2.n.logn)$. When the size of the problem instance (i.e. large values of $n$ and $m$) increases, 
the grain size of the kernel executed by each thread becomes higher which significantly increases the GPU throughput.
For instance, for the problems of the category 200 jobs over 20 machines, the reported speedup of our approach is about $\times$100,48
while the speedup calculated for the multithreaded version is $\times$8,76 which corresponds to an improvement of $\times$11,47. 
Over all the experimented instance categories, the GPU-based B\&B run faster than the multithreaded-based B\&B.

\section{Conclusion and Future Work}
\label{Conclusion}

In this paper, we have revisited the parallel B\&B algorithm for solving permutation-based combinatorial optimization 
problems such as FSP on GPU accelerators. The contributions consist in proposing: (1) a GPU-based parallel design and 
implementation of the parallel bounding model~;~(2) a data access optimization approach to take into account the memory 
constraints of the GPU device. The Flow-Shop scheduling problem has been considered as a case study together with the 
Johnson's lower bound~\cite{SMJohnson_54}, extended in~\cite{JKLenstra_et_al_78} to more than two machines. The proposed 
approaches have been experimented using a Tesla C2050 GPU card on 4 different classes of FSP instances.

In our proposed GPU-based approach, the decomposition and pruning of the sub-problems is performed on CPU and the evaluation 
of their lower bounds (bounding operation) is executed on GPU. Pools of sub-problems are off-loaded from CPU to GPU to be 
evaluated by blocks of threads. After evaluation, the lower bounds are returned back to the CPU. The experimental results 
show that accelerations up to $\times 77$ can be obtained especially for large problem instances and large pools of sub-problems. 
As shown in the reported results the pool size that enables to achieve the best acceleration of the bounding mechanism depends 
strongly on the size of the problem instance being solved. Therefore, this parameter has to be determined at runtime by testing different pool sizes. 

The proposed data access optimization is based on a preliminary analysis of the lower bound function. Such analysis allowed us 
to identify six data structures for which we have proposed a complexity analysis in terms of memory size and access frequency. 
Due to the limited size of the shared memory the matrices do not fit in all together. According to the complexity study, the 
recommendation is to put in the shared memory the Johnson's and the processing time matrices ($JM$ and $PTM$) if they fit in 
together. The other data structures are mapped to the global memory combined with the L1 cache (see \ref{access}). Such recommendation has been confirmed through extensive 
experiments using the Taillard's benchmarks of the Flow-Shop problem and a recent C2050 Tesla GPU card. The optimizations 
obtained with the proposed approaches allowed us to achieve accelerations up to $\times 100$ compared to a single CPU-based B\&B and 
up to $\times 11$ compared to a multi-threaded CPU-based execution.

We are currently investigating the combination of the GPU-based bounding model with the multi-core parallel search tree 
exploration for the design and implementation of a GPU-accelerated multi-core B\&B algorithm. In the near future, we plan 
to extend this work to a cluster of GPU-accelerated multi-core processors. From application point of view, the objective is 
to solve to optimality challenging difficult and unsolved Flow-Shop instances as we did it for one $50 \times 20$ problem instance
using grid computing~\cite{Mezmaz_2007}. Finally, we plan to investigate other lower bound functions to deal with other combinatorial optimization problems.


\begin{thebibliography}{1}
{\scriptsize

\bibitem{BGendron_et_al_94}
B.~Gendron and T.G. Crainic.
\newblock {Parallel Branch-and-Bound Algorithms: Survey and Synthesis}.
\newblock {\em Operations Research}, 42(06):1042--1066, 1994.

\bibitem{Chapman_2007}
B. Chapman, G. Jost, and R. Van Der Pas. 
\newblock {Using OpenMP: portable shared memory parallel programming}.
\newblock { Volume 10. The MIT Press}, 2007.

\bibitem{JKLenstra_et_al_78}
B. J. Lageweg, J. K. Lenstra and A. H. G. Rinnooy Kan.
\newblock {A general bounding scheme for the permutation flow-shop problem}.
\newblock {\em Operations Research}, 26(1):53--67, 1978.

\bibitem{MRGary_et_al_79}
M.R. Garey and D.S. Johnson.
\newblock {\em {Computers and Intractability: A Guide to the Theory of
  NP-Commpleteness}}.
\newblock W. H. Freeman \& Co., New York, NY, 1979.

\bibitem{SMJohnson_54}
S.M. Johnson.
\newblock{Optimal two and three-stage production schedules with setup times included}.
\newblock{Naval Research Logistis Quarterly}, 1:61-68. 1954.

\bibitem{Taillard_Bench}
E. Taillard.
\newblock{Taillard's FSP benchmarks}.
http://mistic.heig-vd.ch/taillard/problemes.dir/ordonnancement.dir/ordonnancement.html.

\bibitem{Tschoke_1995}
S. Tschoke, R. Lubling and B. Monien.
\newblock{Solving the traveling salesman problem with a distributed branch-and-bound algorithm on a 1024 processor network}.
\newblock{In Proc. of $9^{th}$ Intl. Parallel Processing Symposium (IPPS), pp. 182 - 189}, 1995.

\bibitem{Allen_1997}
R. Allen, L. Cinque, S. Tanimoto, L. Shapiro and D. Yasuda.
\newblock{A parallel algorithm for graph matching and its MasPar implementation}.
\newblock{IEEE Transactions on Parallel and Distributed Systems, Vol. 8, No. 5}, 1997.

\bibitem{Casadoa_2008}
L.G. Casadoa, J.A. Martíneza, I. Garcíaa and E.M.T. Hendrixb.
\newblock{Branch-and-Bound interval global optimization on shared memory multiprocessors}.
\newblock{Optimization Methods and Software, Vol. 23, No.5, pp. 689-701}, 2008.

\bibitem{Barreto} 
L. Barreto and M. Bauer.
\newblock{Parallel Branch and Bound Algorithm-A comparison between serial, OpenMP and MPI implementations}. 
\newblock{Journal of Physics: Conference Series, Vol. 256, No.5, pp. 012018}, 2010.

\bibitem{Mezmaz_2007}
M. Mezmaz, N. Melab and E-G. Talbi.
\newblock{A Grid-enabled Branch and Bound Algorithm for Solving Challenging Combinatorial Optimization Problems}.
\newblock{In Proc. of 21th IEEE Intl. Parallel and Distributed Processing Symp. (IPDPS), Long Beach, California, March 26th-30th}, 2007.

\bibitem{Nichols_1996} 
B. Nichols, D. Buttlar, and J.P. Farrell. 
\newblock{Pthreads programming}. 
\newblock{O'Reilly Media}, 1996.

\bibitem{Djamai_2011}
M. Djamai, B. Derbel and N. Melab.
\newblock{Distributed B\&B: A Pure Peer-to-Peer Approach}.
\newblock{In Proc. of IEEE IPDPS'2011, Woks. on Large-Scale Parallel Processing (LSPP), May 16-20, Anchorage (Alaska)}, 2011.

\bibitem{Luong_2012}
T-V. Luong, N. Melab and E-G. Talbi.
\newblock{GPU Computing for Parallel Local Search Metaheuristic Algorithms}.
\newblock{IEEE Transactions on Computers, http://doi.ieeecomputersociety.org/10.1109/TC.2011.206}, 2012.

\bibitem{paulavicius}
R.Paulavi{\v{c}}ius and J. {\v{Z}}ilinskas.
\newblock{Parallel branch and bound algorithm with combination of Lipschitz bounds over multidimensional simplices for multicore computers}.
\newblock{Parallel Scientific Computing and Optimization, Springer, pages 93--102},2009.

\bibitem{sanjuan}
JF. Sanjuan-Estrada, LG. Casado and  I. Garc{\'\i}a.
\newblock{Adaptive parallel interval branch and bound algorithms based on their performance for multicore architectures},
\newblock{The Journal of Supercomputing, Springer, pages 1--9},2011.

\bibitem{NvidiaCudaGuide}
\newblock{NVIDIA CUDA C Programming Best Practices Guide.}
\newblock{http://developer.download.nvidia.com/compute/cuda/2\_3/toolkit/docs/
NVIDIA\_CUDA\_BestPracticesGuide\_2.3.pdf}.

\bibitem{NVIDIA_C2050} http://www.nvidia.com/docs/IO/43395/NV\_DS\_Tesla\_C2050\_C2070\_jul10\_lores.pdf

\bibitem{NVIDIA_C2050_2} http://en.wikipedia.org/wiki/Comparison\_of\_Nvidia\_graphics\_processing\_units

\bibitem{INTEL_I7} http://download.intel.com/support/processors/corei7/sb/core\_i7-900\_d.pdf

\bibitem{INTEL_I7_2}http://ark.intel.com/products/47933/Intel-Core-i7-970-Processor-\%2812M-Cache-3\_20-GHz-4\_80-GTs-Intel-QPI\%29

\bibitem{TOP500} http://www.top500.org/
}

\end{thebibliography}
\end{document}